\documentclass{PoS}

\title{HYDJET++ heavy ion event generator and its applications for RHIC and LHC}

\ShortTitle{HYDJET++ heavy ion event generator and its applications for RHIC and LHC}

\author{\speaker{I.P. Lokhtin}, L.V. Malinina, S.V. Petrushanko, 
A.M. Snigirev\\
         D.V. Skobeltsyn Institute of Nuclear Physics, M.V. Lomonosov Moscow
	State University, Moscow, Russia \\
        E-mail: \email{Igor@Lokhtin@cern.ch}}

\author{I. Arsene\thanks{On leave from the Institute for Space Sciences, 
Bucharest, Romania}\\
       The Department of Physics, University of Oslo, Norway\\}

\author{K. Tywoniuk\\ Departamento de 
F{\'\i}sica de Part{\'\i}culas, Universidad de Santiago de Compostela, 
Santiago de Compostela, Spain}

\abstract{The heavy ion event generator HYDJET++ is presented. HYDJET++
simulates  relativistic heavy ion AA collisions as a superposition 
of the soft, hydro-type state and the hard state resulting from multi-parton 
fragmentation. This model is the development and continuation of HYDJET event 
generator. The hard parts of HYDJET and HYDJET++ are identical. The soft part 
of HYDJET++ contains the following important additional features as compared with HYDJET: 
resonance decays and more detailed treatment of thermal and chemical 
freeze-out hypersurfaces. HYDJET++ is capable of reproducing the bulk 
properties of heavy ion collisions at RHIC (hadron spectra and ratios, radial 
and elliptic flow, femtoscopic momentum correlations), as well as high-p$_T$ 
hadron spectra. Some applications of HYDJET++ at LHC are discussed.}  

\FullConference{4th international workshop High-pT physics at LHC 09\\

        February 4-7, 2009\\

        Prague, Czech Republic}

\begin{document}
\def\la{\mathrel{\mathpalette\fun <}}
\def\ga{\mathrel{\mathpalette\fun >}}
\def\fun#1#2{\lower3.6pt\vbox{\baselineskip0pt\lineskip.9pt
\ialign{$\mathsurround=0pt#1\hfil##\hfil$\crcr#2\crcr\sim\crcr}}} 
\newcommand{\Pom}{{\hspace{-0.1em}I\hspace{-0.25em}P}}

\section{Introduction}
One of the basic tasks of modern high energy physics is the study of the 
fundamental theory of strong interaction (Quantum Chromodynamics, QCD) in 
new, unexplored extreme regimes of super-high densities and temperatures through  
the investigation of the properties of multi-particle systems produced in high-energy nuclear 
collisions~\cite{d'Enterria:2006su,BraunMunzinger:2007zz}. Ongoing and 
future experimental heavy ion studies require the development of new Monte-Carlo 
(MC) event generators and improvement of existing ones. Especially for experiments 
at the CERN Large Hadron Collider (LHC), because of very high parton 
and hadron multiplicities, one needs fast (but realistic) MC tools for heavy 
ion event simulations~\cite{alice1,alice2,cms,Abreu:2007kv}. A realistic MC 
event generator should include a maximum possible number of observable physical 
effects which are important to determine the event topology: from the bulk 
properties of soft hadroproduction (domain of low transverse momenta $p_T \la 
1$GeV$/c$) such as collective flows, to hard multi-parton production in hot and dense 
QCD-matter, which reveals itself in the spectra of high-$p_T$ particles and 
hadronic jets. However, in most of the available MC heavy ion event generators, 
the simultaneous treatment of collective flow effects for soft hadroproduction 
and hard multi-parton in-medium production is absent. 

HYDJET++ event generator~\cite{Lokhtin:2008xi,hydjet++} includes detailed 
treatment of soft hadroproduction as well as hard multi-parton production, and 
takes into account medium-induced parton rescattering and energy loss. The 
heavy ion event in HYDJET++ is the superposition of two independent components: 
the soft, hydro-type state and the hard state resulting from multi-parton 
fragmentation. HYDJET++ model is the development and continuation of HYDJET 
event generator~\cite{Lokhtin:2005px,Lokhtin:2007ga,Lokhtin:2008wg,hydjet}, and it  
contains the important additional features for the soft component: resonance decays and 
more detailed treatment of thermal and chemical freeze-out 
hypersurfaces~\cite{Amelin:2006qe,Amelin:2007ic}. The main program HYDJET++ is 
written in the object-oriented C++ language under the ROOT environment~\cite{root}.

\section{Physics model and simulation procedure}

The soft and hard components in HYDJET++ are treated independently. When the 
generation of soft and hard components in each event at given $b$ is completed, the 
event record (information about coordinates and momenta of primordial particles, 
decay products of unstable particles and stable particles) is formed as the junction 
of these two independent event outputs. 

The details on physics model and simulation procedure of HYDJET++ can be found in the 
corresponding manual~~\cite{Lokhtin:2008xi}. The main features of HYDJET++ model are 
listed only very briefly in this section.

\subsection{Hard multi-jet production} 

The model for the hard multi-parton part of HYDJET++ event is the same as that 
for HYDJET event generator, and it based on PYQUEN partonic energy loss 
model~\cite{Lokhtin:2005px,Lokhtin:2007ga,Lokhtin:2008wg}. The approach to the 
description of multiple scattering 
of hard partons in the dense QCD-matter (such as quark-gluon plasma) is based on the 
accumulative energy loss via  the gluon radiation being associated with each parton 
scattering in the expanding quark-gluon fluid and includes the interference effect 
(for the emission of gluons with a finite formation time) using the modified radiation 
spectrum $dE/dl$ as a function of decreasing temperature $T$. The model takes into
account radiative and collisional energy loss of hard partons in longitudinally
expanding quark-gluon fluid, as well as realistic nuclear geometry.  

The Fortran routine for single hard nucleon-nucleon 
sub-collision PYQUEN~\cite{pyquen} was constructed as a modification of the jet event 
obtained with the generator of hadron-hadron interactions PYTHIA$\_$6.4~\cite{pythia}. 
The event-by-event simulation procedure in PYQUEN includes {\it 1)} generation of 
initial parton spectra with PYTHIA and production vertexes at given impact parameter; 
{\it 2)} rescattering-by-rescattering simulation of the parton path in a dense zone 
and its radiative and collisional energy loss; {\it 3)} final hadronization according 
to the Lund string model for hard partons and in-medium emitted gluons. Then the 
PYQUEN multi-jets generated according to the binomial distribution are included in the 
hard part of the event. The mean number of jets produced in an AA event is the 
product of the number of binary NN sub-collisions at a given impact parameter and the 
integral cross section of the hard process in $NN$ collisions with the minimum  
transverse momentum transfer $p_T^{\rm min}$. In order to take into account the 
effect of nuclear shadowing on parton distribution functions, the impact parameter 
dependent parameterization obtained in the framework of Glauber-Gribov 
theory~\cite{Tywoniuk:2007xy} is used. 

Note that some different approaches for MC treatment of partonic energy loss, such as
codes YaJEM~\cite{Renk-LHC09}, JEWEL~\cite{Zapp-LHC09} and
Q-PYTHIA~\cite{Mendez-LHC09} have been presented during this Workshop.

\subsection{Soft ``thermal'' hadron production}

The soft part of HYDJET++ event is the ``thermal'' hadronic state generated on the 
chemical and thermal freeze-out hypersurfaces obtained from the parameterization 
of relativistic hydrodynamics with preset freeze-out conditions (the adapted C++ code 
FAST MC~\cite{Amelin:2006qe,Amelin:2007ic}). Hadron multiplicities are calculated 
using the effective thermal volume approximation and Poisson multiplicity distribution 
around its mean value, which is supposed to be proportional to the number of 
participating nucleons at a given impact parameter of AA collision. The fast soft 
hadron simulation procedure includes {\it 1)} generation of the 4-momentum of a hadron 
in the rest frame of a liquid element in accordance with the equilibrium distribution 
function; {\it 2)} generation of the spatial position of a liquid element and its 
local 4-velocity in accordance with phase space and the character of motion of the 
fluid; {\it 3)} the standard von Neumann rejection/acceptance procedure to account 
for the difference between the true and generated probabilities; {\it 4)} boost of 
the hadron 4-momentum in the center mass frame of the event; {\it 5)} the two- 
and three-body decays of resonances with branching ratios taken from the SHARE 
particle decay table~\cite{share}. The high generation speed in HYDJET++ is achieved 
due to almost 100\% generation efficiency of the ``soft'' part because of the  
nearly uniform residual invariant weights which appear in the freeze-out 
momentum and coordinate simulation. 

Let us indicate some physical restrictions of the model. HYDJET++ is only 
applicable for symmetric AA collisions of heavy (A $\ga 40$) ions at high 
energies ($\sqrt{s} \ga 10$ GeV). Since the hydro-type approximation for heavy 
ion collisions is considered to be valid for central and semi-central collisions, 
the results obtained for very peripheral collisions (with impact parameter of 
the order of two nucleus radii, $b \sim 2 R_A$) may be not adequate. Nor do we  
expect a correct event description in the region of very forward rapidities, 
where the other mechanisms of particle production, apart from hydro-flow and jets, 
may be important. 

\section{HYDJET++ software structure} 

\begin{figure}
\begin{center}
\includegraphics[angle=270,width=12cm]{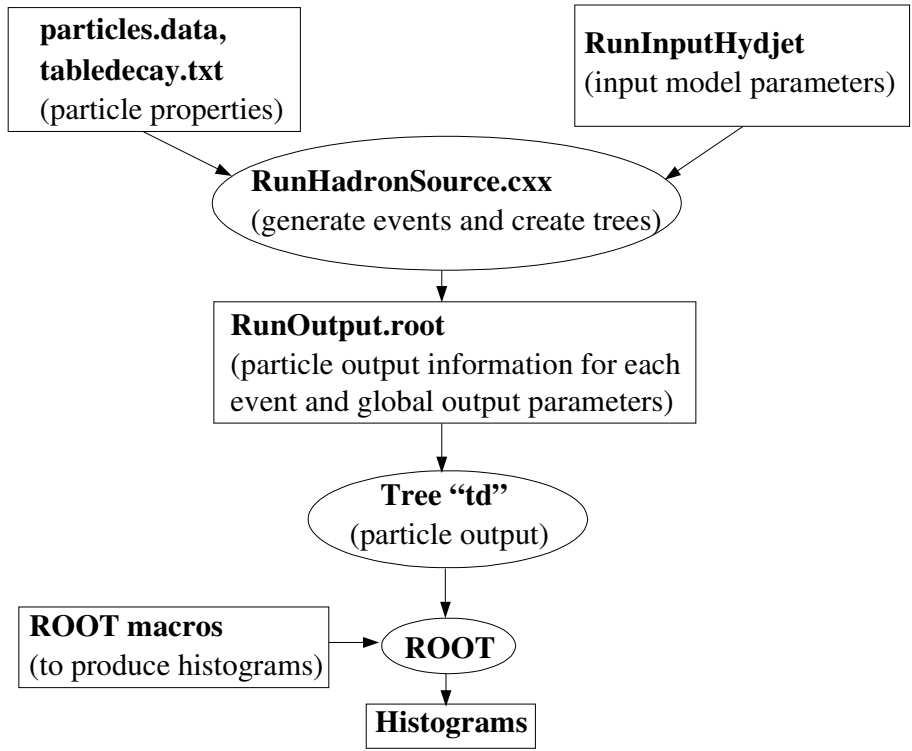}
\end{center}
\caption{The block structure of HYDJET++.}
\label{fig_hydjet_block}
\end{figure}

The basic frameworks of HYDJET++ are preset by the object-oriented C++ language and 
the ROOT environment~\cite{root}. There is also the Fortran-written part~\cite{hydjet} 
which is included in the generator structure as a separate directory. 
The block structure of HYDJET++ is shown in Figure \ref{fig_hydjet_block}. The main 
program elements are particle data files, input and output files, C++ and Fortran 
routines.  

The information regarding the particle species included in the 
HYDJET++ event is stored in the files \verb*|particles.data| and 
\verb*|tabledecay.txt|. The \verb*|particles.data| file contains the definition (PDG 
code) and physical properties (mass, decay width, spin, isospin, valence quark 
composition) of 360 stable hadrons and resonances. The \verb*|tabledecay.txt| file 
contains decay channels and branching ratios. The structure of these files is the 
same as that in SHARE particle data table~\cite{share} and in event generator 
THERMINATOR~\cite{therminator}. 

Run of HYDJET++ is controlled by the file \verb*|RunInputHydjet| for different 
type of input parameters. The input file contains 7 input parameters (number of
events to generate, beam c.m.s. energy per nucleon pair in GeV, atomic weight of 
nuclei and parameters to specify the type of centrality selection) and 18 free
model parameters, which can be varied by the user from their default values (chemical
potentials, chemical and thermal freeze-out temperatures, space-time scales of soft 
hadron emission, maximal longitudinal and transverse flow rapidities, momentum and
coordinate azimuthal anisotropy parameters, PYQUEN energy loss model parameters). A 
number of important PYTHIA parameters also may be changed/specified in 
\verb*|RunInputHydjet| file. There are also 10 flags to specify different physical 
model scenarios for soft and/or hard components. In particular, the activation of 
some flags allows one to calculate a few model parameters, and so to reduce the 
number of independent free parameters (down to 12 at the minimum). Two additional files 
with the optimized parameters for Au+Au collisions at $\sqrt{s}=200 A$ GeV 
(\verb*|RunInputHydjetRHIC200|) and for Pb+Pb collisions at $\sqrt{s}=5500 A$ GeV 
(\verb*|RunInputHydjetLHC5500|) are available in the distribution package. The default 
parameters for \verb*|RunInputHydjetRHIC200| were obtained by fitting RHIC data to 
various physical observables. The default parameters for \\ \verb*|RunInputHydjetLHC5500| 
represent our rough extrapolation from RHIC to LHC energy. 

The program output is directed to the ROOT file \verb*|RunOutput.root|. The output 
file contains a tree
named \verb*|td|, which keeps the entire event record including primary particles
and decay products with their coordinates and momenta information. Each decay
product contains the unique index of its parent particle so that the entire event 
history may be obtained. Beside particle information, the output file contains also 
a number of global output parameters for each event (generated value of impact
parameter, total inelastic and hard scattering cross sections, hadron event 
multiplicities of hard and soft components, numbers of binary NN sub-collisions and 
nucleons-participants). 

HYDJET++ includes 17 C++ source files (4 main modules and 13 service
modules placed in the main directory) and 3 Fortran files (placed in the separate
directory). The size of the distribution package is 3.5 MBytes directory and 800 
kBytes compressed archive (without ROOT libraries). The generation of 
100 central (0$-$5\%) Au+Au events at $\sqrt{s} = 200 A$ GeV (Pb+Pb events at 
$\sqrt{s} = 5500 A$ GeV) with default input parameters takes about 7 (85) minutes on 
a PC 64 bit Intel Core Duo CPU @ 3 GHz with 8 GB of RAM memory under Red Hat 
Enterprise. Then the output file created by the code in ROOT tree format will require 
$40$ ($190$) MBytes of the disk space. 

\section{Validation of HYDJET++ with experimental RHIC data}  

It was demonstrated in~\cite{Amelin:2006qe,Amelin:2007ic} that FAST MC model can 
describe well the bulk properties of hadronic state created in Au+Au collisions 
at RHIC at $\sqrt{s}=200 A$ GeV (such as particle number ratios, low-p$_T$ spectra, 
elliptic flow coefficients $v_2(p_T, b)$, femtoscopic correlations in 
central collisions), while HYDJET model is 
capable of reproducing the main features of jet quenching pattern at RHIC 
(high-$p_T$ hadron spectra and the suppression of azimuthal back-to-back 
correlations)~\cite{Lokhtin:2005px}. Since soft and hard hadronic states in
HYDJET++ are simulated independently, a good description of hadroproduction at 
RHIC in a wide kinematic range can be achieved, moreover a number of 
improvements in FAST MC and HYDJET have been done as compared to earlier 
versions. A number of input parameters of the model can be fixed from fitting 
the RHIC data to various physical observables. 

\begin{figure}
\begin{center}
\includegraphics[width=14cm]{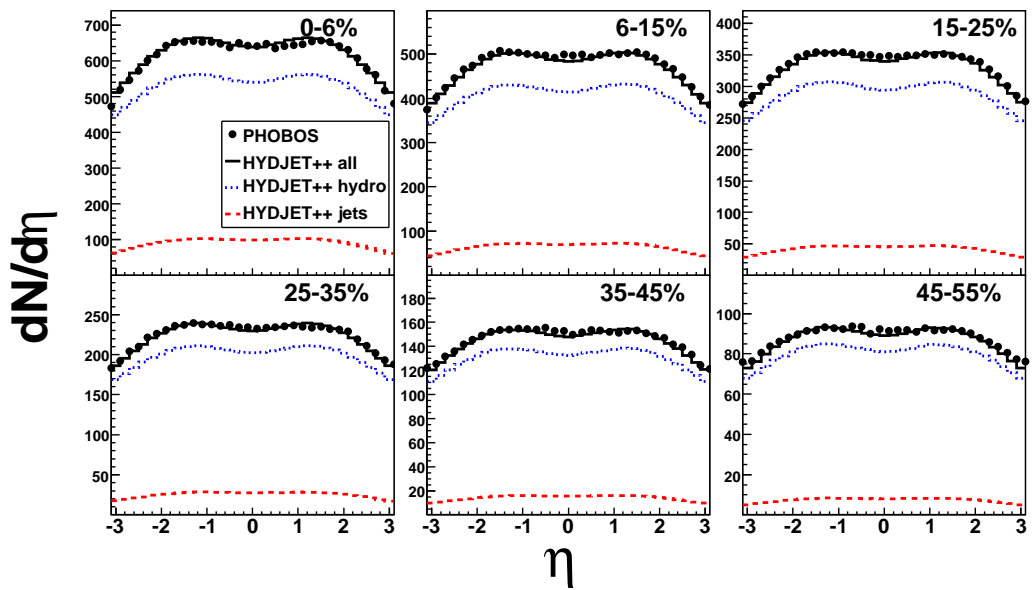}
\end{center}
\caption{The pseudorapidity distribution of charged hadrons in Au+Au 
collisions at $\sqrt{s}=200 A$ GeV for six centrality sets. The points are 
PHOBOS data~\cite{Back:2002wb}, histograms are the HYDJET++ calculations 
(solid -- total, dotted -- hydro part, dashed -- jet part).}
\label{fig_eta_all}
\end{figure}

\begin{figure}
\begin{center}
\includegraphics[width=14cm]{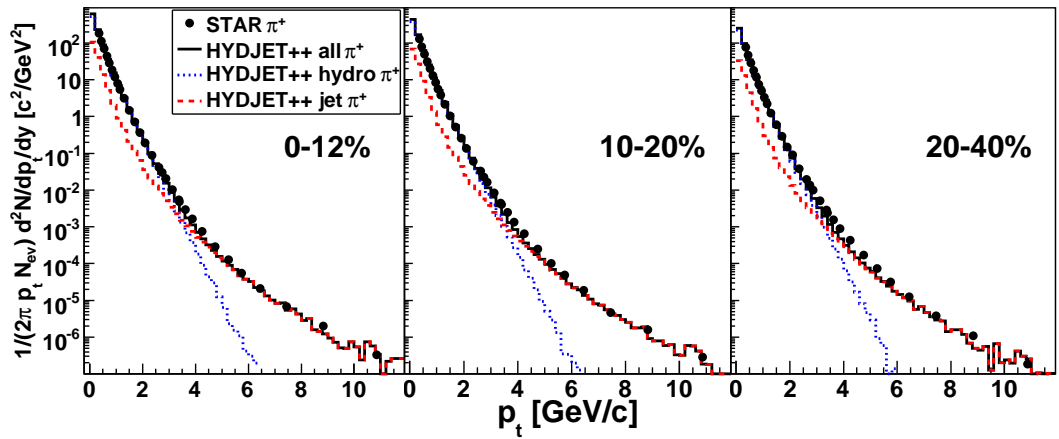}
\end{center}
\caption{The transverse momentum distribution of positively charged pions in 
Au+Au collisions at $\sqrt{s}=200 A$ GeV for three centrality sets. The points 
are STAR data~\cite{Abelev:2006jr}, histograms are the HYDJET++ calculations 
(solid -- total, dotted -- hydro part, dashed -- jet part).}
\label{fig_hpt_all}
\end{figure}

\begin{figure}
\begin{center}
\includegraphics[width=14cm]{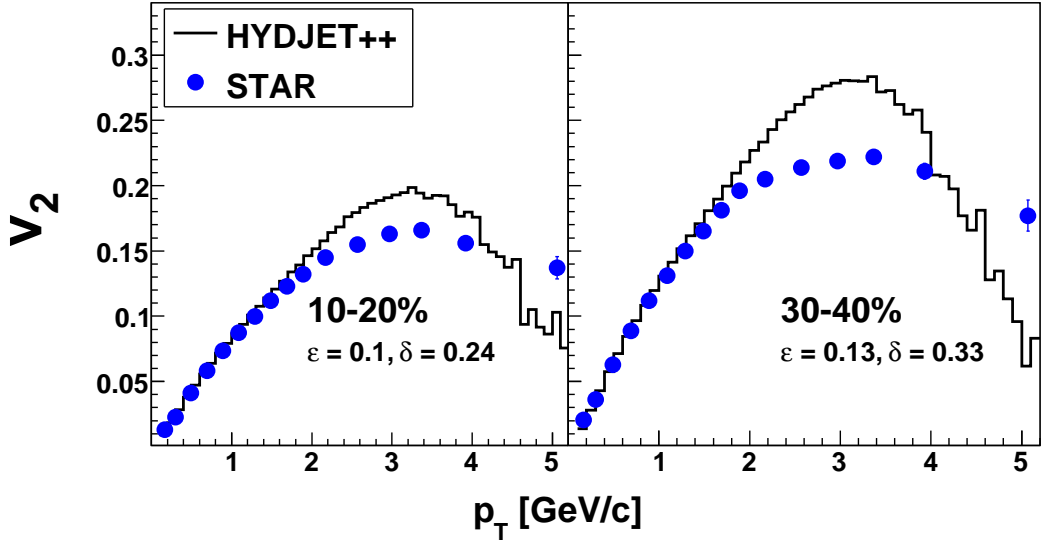}
\end{center}
\caption{The transverse momentum dependence of the elliptic flow  
coefficient $v_2$ of charged hadrons in Au+Au collisions at $\sqrt{s}=200 A$ 
GeV for two centrality sets. The points are STAR data~\cite{Adams:2004bi}, 
histograms are the HYDJET++ calculations.}
\label{fig_v2_all}
\end{figure}

\begin{enumerate}  

\item {\bf Ratio of hadron abundances.} It is well known that the particle
abundances in heavy ion collisions in a wide energy range can be reasonable well
described within statistical models based on the assumption that the produced
hadronic matter reaches thermal and chemical equilibrium. The thermodynamical
potentials $\widetilde{\mu_{B}}=0.0285$ GeV, $\widetilde{\mu_{S}}=0.007$ GeV, 
$\widetilde{\mu_{Q}}=-0.001$, the strangeness suppression factor $\gamma_s=1$, 
and the chemical freeze-out temperature  $T^{\rm ch}=0.165$ GeV have been fixed 
in~\cite{Amelin:2006qe} from fitting the RHIC data to  
various particle ratios near mid-rapidity in central Au+Au collisions at 
$\sqrt{s}=200 A$ GeV ($\pi^-/\pi^+$, $\bar{p}/\pi^-$, $K^-/K^+$, $K^-/\pi^-$, 
$\bar{p}/p$, $\bar{\Lambda}/\Lambda$, $\bar{\Lambda}/\Lambda$, $\bar{\Xi}/\Xi$, 
$\phi/K^-$, $\Lambda/p$, $\Xi^-/\pi^-$). 

\item {\bf Low-$p_T$ hadron spectra.} Transverse momentum $p_T$ and transverse 
mass $m_T$ hadron spectra ($\pi^+$, $K^+$ and $p$ with $m_T<0.7$ GeV/$c^2$) near  
mid-rapidity at different centralities of Au+Au collisions at $\sqrt{s}=200 A$ 
GeV were analyzed in~\cite{Amelin:2007ic}. The slopes of these
spectra allow the thermal freeze-out temperature  $T^{\rm th}=0.1$ GeV and 
the maximal transverse flow rapidity in central collisions 
$\rho_u^{\rm max}(b=0)=1.1$ to be fixed.  

\item {\bf Femtoscopic correlations.} Because of the effects of quantum 
statistics and final state interactions, the momentum (HBT) correlation functions 
of two or more particles at small relative momenta in 
their c.m.s. are sensitive to the space-time characteristics of 
the production process on the level of $fm$. The space-time parameters of thermal 
freeze-out region in central Au+Au collisions at $\sqrt{s}=200 A$ GeV have been 
fixed in~\cite{Amelin:2007ic} by means of fitting the three-dimensional correlation 
functions measured for $\pi^+\pi^+$ pairs and extracting the correlation radii 
$R_{\rm side}$, $R_{\rm out}$ and $R_{\rm long}$: $\tau_f(b=0)=8$ fm/$c$, 
$\Delta \tau_f(b=0)=2$ fm/$c$, $R_f(b=0)=10$ fm. 

\item {\bf Pseudorapidity hadron spectra.} The PHOBOS data on $\eta$-spectra of 
charged hadrons~\cite{Back:2002wb} at different centralities of Au+Au collisions 
at $\sqrt{s}=200 A$ GeV have been analyzed to fix the particle densities in the 
mid-rapidity region and the maximum longitudinal flow rapidity 
$\eta_{\rm max}=3.3$ (Fig. \ref{fig_eta_all}). Since mean ``soft'' and ``hard''
hadron multiplicities depend on the centrality in different ways (they are 
roughly proportional to $\overline{N_{\rm part}(b)}$ and 
$\overline{N_{\rm bin}(b)}$ respectively), the relative contribution of soft 
and hard parts to the total event multiplicity can be fixed through the 
centrality dependence of $dN/d\eta$. The corresponding contributions from 
hydro- and jet-parts are determined by the input parameters 
$\mu_{\pi}^{\rm eff~th}=0.053$ GeV and $p_T^{\rm min}=3.4$ GeV/$c$ respectively.

\item {\bf High-$p_T$ hadron spectra.} High transverse momentum hadron 
spectra ($p_T \ga 2-4$ GeV/$c$) are sensitive to parton production and 
jet quenching effects. Thus fitting the measured high-$p_T$ tail allows 
the extraction of PYQUEN energy loss model parameters. We assume 
the QGP formation time $\tau_0=0.4$ fm/$c$ and the number of active quark 
flavours $N_f=2$. Then the reasonable 
fit of STAR data on high-$p_T$ spectra of charged pions at different 
centralities of Au+Au collisions at $\sqrt{s}=200 A$ GeV~\cite{Abelev:2006jr} is
obtained with the initial QGP temperature $T_0=0.3$ GeV (Fig. \ref{fig_hpt_all}).  

\item {\bf Elliptic flow.} The elliptic flow coefficient $v_2$ (which is
determined as the second-order Fourier coefficient in the hadron distribution
over the azimuthal angle $\varphi$ relative to the reaction plane angle
$\psi_R$, so that $v_2 \equiv \left< \cos{2(\varphi-\psi_R)} \right>$) is an
important signature of the physics dynamics at early stages of non-central 
heavy ion collisions. According to the typical hydrodynamic scenario, the
values $v_2(p_T)$ at low-$p_T$ ($\la 2$ GeV/$c$) are determined mainly by the 
internal pressure gradients of an expanding fireball during the initial high 
density phase of the reaction (and it is sensitive to the momentum and azimuthal 
anisotropy parameters $\delta$ and $\epsilon$ in the frameworks of HYDJET++), while 
elliptic flow at high-$p_T$ is generated in HYDJET++ (as well as in other jet
quenching models) due to the partonic energy loss in an azimuthally asymmetric 
volume of QGP. Figure \ref{fig_v2_all} shows the measured by the STAR 
Collaboration transverse momentum dependence of the elliptic flow coefficient 
$v_2$ of charged hadrons in Au+Au collisions at $\sqrt{s}=200 A$ GeV for two  
centrality sets~\cite{Adams:2004bi}. The values of $\delta$ and $\epsilon$ are 
estimated for each centrality. Note that the choice 
of these parameters does not affect any azimuthally integrated physics 
observables (such as hadron multiplicities, $\eta$- and $p_T$-spectra, etc.), 
but only their differential azimuthal dependences.

\end{enumerate}

\section{Simulations with HYDJET++ at LHC}

The heavy ion collision energy at LHC a factor of $30$ larger then that at  
RHIC, thereby allows one to probe new 
frontiers of super-high temperature and (almost) net-baryon free 
QCD. The emphasis of the LHC heavy ion data 
analysis (at $\sqrt{s}=5.5$ TeV per nucleon pair for lead beams) will be on the 
perturbative, or hard probes of the QGP (quarkonia, jets, photons, high-p$_T$ 
hadrons) as well as on the global event properties, or soft probes (collective 
radial and elliptic flow effects, hadron multiplicity, transverse energy 
densities and femtoscopic momentum  correlations). It is expected 
that at LHC energies the role of hard and semi-hard particle production will be 
significant even for the bulk properties of created matter. HYDJET++ seems to be
an effective simulation tool to analyze the influence of in-medium jet 
fragmentation on various physical observables. 

\begin{figure}
\begin{center}
\includegraphics[width=13.5cm]{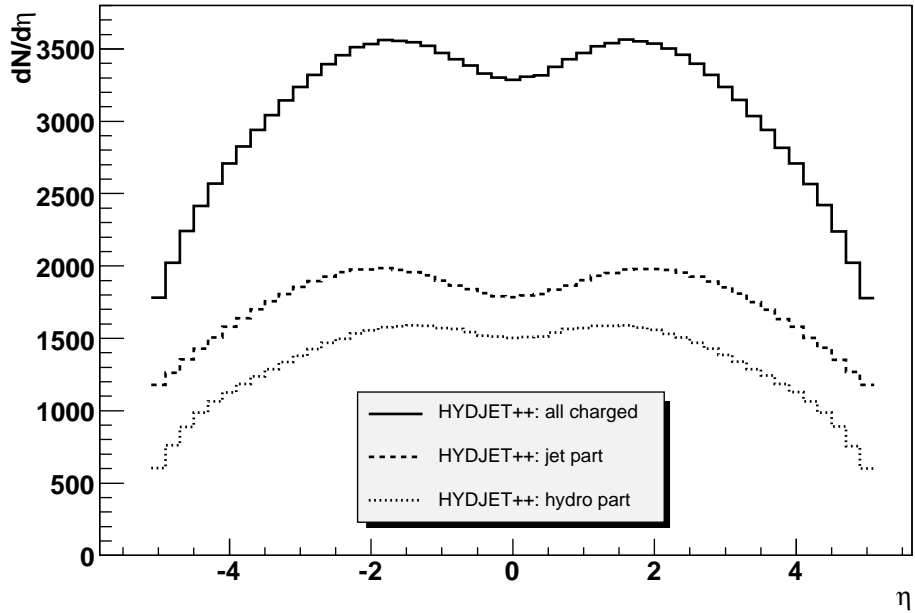}
\end{center}
\caption{The pseudorapidity distribution of charged hadrons in $5$\% most
central Pb+Pb collisions at $\sqrt{s}=5500 A$ GeV (solid -- total, dotted -- 
hydro part, dashed -- jet part), $p_T^{\rm min}=7$ GeV/$c$.}
\label{eta7-lhc}
\end{figure}

\begin{figure}
\begin{center}
\includegraphics[width=13.5cm]{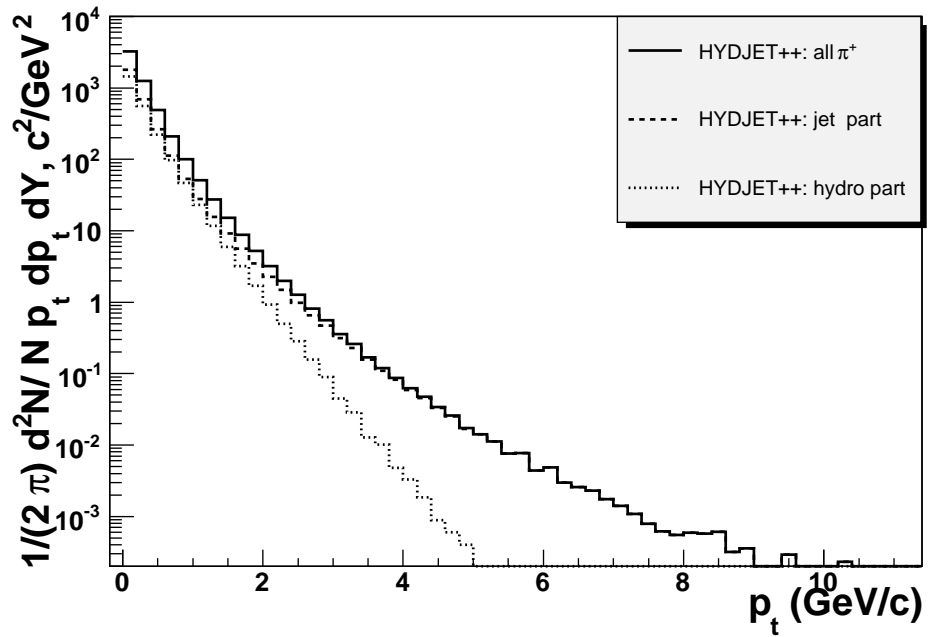}
\end{center}
\caption{The transverse momentum distribution of pions in $5$\% most
central Pb+Pb collisions at $\sqrt{s}=5500 A$ GeV (solid -- total, dotted -- 
hydro part, dashed -- jet part), $p_T^{\rm min}=7$ GeV/$c$.}
\label{pt7-lhc}
\end{figure}

Figures \ref{eta7-lhc} and 
\ref{pt7-lhc} show the pseudorapidity distribution of charged hadrons and 
transverse momentum distribution of pions respectively obtained with HYDJET++ 
default settings (in particular, $p_T^{\rm min}=7$ GeV/$c$) for $5$\% most
central Pb+Pb events. The estimated contribution of hard component to the total event 
multiplicity is on the level $\sim 55$\% here, what is much larger than as 
compared with RHIC ($\sim 15$\%, see Fig. \ref{fig_eta_all}). Of course, this
number is very sensitive to the parameter $p_T^{\rm min}$ --- minimal 
$p_T$ of ``non-thermalized'' parton-parton hard scatterings. For example,
increasing the value $p_T^{\rm min}$ up to $10$ GeV/$c$ results in decreasing
this contribution down to $\sim 25$\%. 

Some applications of HYDJET++ for high-p$_T$ studies at the LHC have been 
presented during this Workshop~\cite{eyyubova-LHC09,milosevic-LHC09}.
In this paper we discuss one another striking example: the influence of jets 
on femtoscopic momentum correlations (HBT-radii)~\cite{Paic:2004zj}. Since 
HYDJET++ specifies the space-time structure of a hadron emission source (for 
soft and for hard components as well), the momentum correlation function can 
be introduced by the special weighting procedure~\cite{pod83, Pratt84}. Knowing 
the information on final particle four-momenta $p_i$ and 
four-coordinates $x_i$ of the emission points allows one to calculate the 
correlation function with the help of the weight procedure, assigning a weight
to a given particle combination accounting for the effects of quantum 
statistics:
\begin{equation}
      w = 1 + \cos(q \cdot \Delta x),
\label{cf4}
\end{equation}
where  $q = p_1 -p_2 $ and $\Delta x = x_1 - x_2$. Then the correlation 
function is defined as a ratio of the weighted histogram of the pair kinematic
variables to the unweighted one. The corresponding correlation widths are 
parameterized in terms of the Gaussian correlation radii $R_i$,
\begin{equation}
CF(p_{1},p_{2})= 1+\lambda\exp(-R_\mathrm{out}^2q_\mathrm{out}^2
-R_\mathrm{side}^2q_\mathrm{side}^2
-R_\mathrm{long}^2q_\mathrm{long}^2
-2R_\mathrm{out,long}^2q_\mathrm{out}q_\mathrm{long}) \label{cf3}~,
\end{equation}
where $\mathbf{q}=(q_\mathrm{out},q_\mathrm{side},q_\mathrm{long})$ is 
the relative three-momentum vector of two identical particles and $\lambda$ 
is the correlation strength. The {\it out} and {\it side} denote the 
transverse, with respect to the reaction axis, components of the vector 
${\bf q}$; the {out} direction is parallel to the transverse component of the 
pair three--momentum. 

\begin{figure}
\begin{center}
\includegraphics[width=14cm]{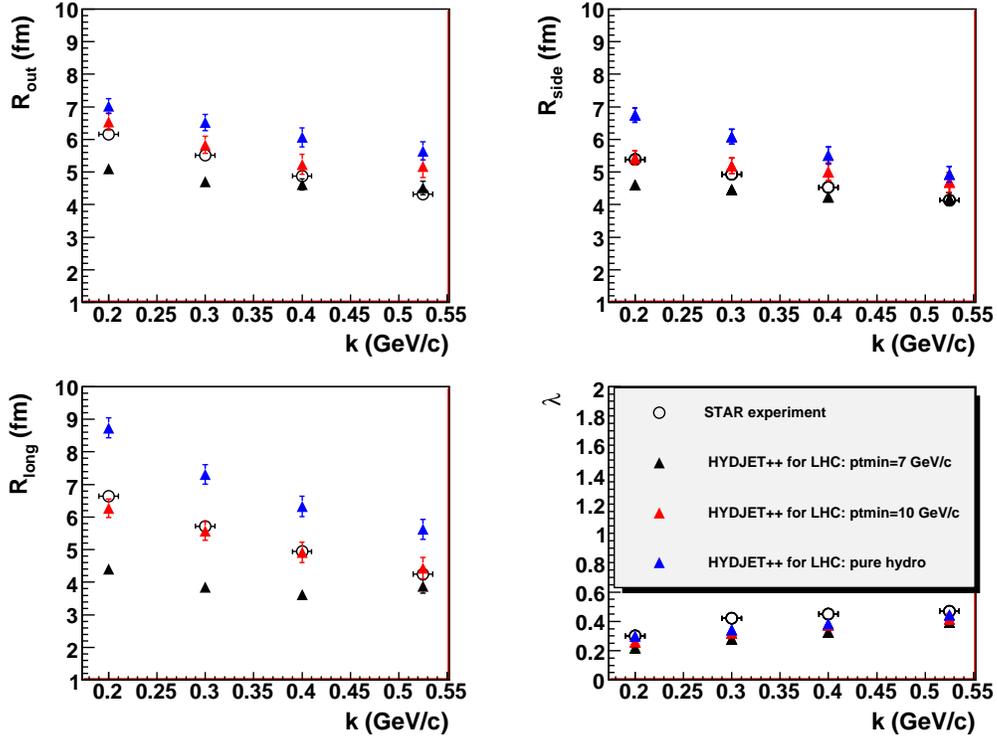}
\end{center}
\caption{The $\pi^\pm \pi^\pm$ correlation radii and the strength parameter 
$\lambda$ at mid-rapidity as the function of relative pion momentum $k$ in 
$5$\% most central Pb+Pb collisions at $\sqrt{s}=5500 A$ GeV 
(black triangles --  $p_T^{\rm min}=7$ GeV/$c$, red triangles -- 
$p_T^{\rm min}=10$ GeV/$c$, blue triangles -- pure hydro). 
The open circles are STAR data~\cite{CF_STAR}.}
\label{hbt-lhc}
\end{figure}

The correlation functions of two identical charged pions have been calculated 
for $5$\% most central Pb+Pb events at two values of $p_T^{\rm min}$: 
$7$ GeV/$c$ and $10$ GeV/$c$. Figure \ref{hbt-lhc} shows the corresponding
values of fitted correlation radii and strength parameter as a function of
relative pion momentum in comparison with those measured by STAR 
collaboration~\cite{CF_STAR}. One can see that the 
``pure hydro'' scenario results in some increasing of correlation radii at LHC
in comparison with RHIC, as it could be naively expected from the appropriate 
extrapolation of the volume parameters~\cite{Abreu:2007kv}. However increasing 
the contribution of hard component to the total event multiplicity results in 
decreasing the correlation radii due to the fact that ``jet-induced'' 
hadrons are emitted on much shorter space-time scales than ``thermal'' 
particles. HYDJET++ predicts that the correlation radii at LHC become 
comparable with those at RHIC for $\sim 25$\% contribution of hard component 
($p_T^{\rm min}=10$ GeV/$c$), and may be even less than at RHIC for larger
contribution of hard component ($p_T^{\rm min}<10$ GeV/$c$). On the other hand, 
the influence of hard component on correlation radii at RHIC was found to be 
negligible. Thus the observation of reducing the correlation radii as one moves 
from RHIC to LHC could manifest the strong influence of in-medium jet 
fragmentation on the bulk event topology. 

\section{Summary}
Among other heavy ion event generators, HYDJET++ focusses on the 
detailed simulation of jet quenching effect basing on the partonic energy loss 
model PYQUEN, and also reproducing the main features of nuclear collective 
dynamics by fast (but realistic) way. The final hadron state in HYDJET++ 
represents the superposition of two independent components: hard multi-parton 
fragmentation and soft hydro-type part. The main program is written in the 
object-oriented C++ language under the ROOT environment. This model is the 
development and continuation of HYDJET event generator. The hard part of 
HYDJET++ is identical to the hard part of Fortran-written HYDJET and it is 
included in the generator structure as a separate directory. The soft part of 
HYDJET++ is the ``thermal'' hadronic state generated on the chemical and 
thermal freeze-out hypersurfaces obtained from the parameterization of 
relativistic hydrodynamics with preset freeze-out conditions. It contains the 
important additional features as compared with HYDJET: resonance decays and 
more detailed treatment of thermal and chemical freeze-out hypersurfaces. 
HYDJET++ is capable of reproducing the bulk properties of heavy ion collisions 
at RHIC (hadron spectra and ratios, radial and elliptic flow, femtoscopic 
momentum correlations), as well as high-p$_T$ hadron spectra. 

HYDJET++ is an effective simulation tool to analyze the influence of 
in-medium jet fragmentation on various physical observables at the LHC. 
In particular, the spectacular prediction of HYDJET++ is reducing the 
femtoscopic correlation radii in heavy ion collisions as one moves from RHIC 
to LHC due to the significant contribution of (semi-)hard component to the 
space-time structure of the hadron emission source. 

\section{Acknowledgments} I.L. wishes to express the gratitude to the organizers of 
the Workshop ``High-pT Physics at LHC'' for the warm welcome and the hospitality. 
This work was supported by Russian Foundation for Basic Research 
(grants No 08-02-91001 and No 08-02-92496), Grants of President of Russian Federation 
(No 107.2008.2 and No 1456.2008.2) and Dynasty Foundation.


\newpage

\begin{thebibliography}{99}
 \bibitem{d'Enterria:2006su} D.~d'Enterria, \emph{J. Phys.} {\bf G 34} (2007) 
S53. 
\bibitem{BraunMunzinger:2007zz} P.~Braun-Munzinger and J.~Stachel, \emph{Nature}
 {\bf 448} (2007) 302. 
\bibitem{alice1} F.~Carminati et al. (ALICE Collaboration), 
\emph{J. Phys.} {\bf G 30} (2004) 1517.
\bibitem{alice2} B.~Alessandro et al. (ALICE Collaboration), 
\emph{J. Phys.} {\bf G 32} (2006) 1295.
\bibitem{cms} D.~d'Enterria et al. (CMS Collaboration), 
\emph{J. Phys.} {\bf G 34} (2007) 2307.
\bibitem{Abreu:2007kv} N.~Armesto (ed.) et al., \emph{J. Phys.} {\bf G 35} (2008)
054001. 
\bibitem{Lokhtin:2008xi} I.P.~Lokhtin, L.V.~Malinina, S.V.~Petrushanko, A.M.~Snigirev, 
I.~Arsene and K.~Tywoniuk, \emph{Comput. Phys. Commun.}, in press, 
{\tt arXiv:0809.2708}. 
\bibitem{hydjet++} http://cern.ch/lokhtin/hydjet++~.
\bibitem{Lokhtin:2005px} I.P.~Lokhtin and A.M.~Snigirev, \emph{Eur. Phys. J.} 
{\bf C 45} (2006) 211. 
\bibitem{Lokhtin:2007ga}  I.P.~Lokhtin, S.V.~Petrushanko, A.M.~Snigirev and 
C.Yu.~Teplov,\pos{PoS (LHC07) 003} [{\tt arXiv:0809.2708}].  
\bibitem{Lokhtin:2008wg} I.P.~Lokhtin, L.V.~Malinina, S.V.~Petrushanko, A.M.~Snigirev, 
I.~Arsene and K.~Tywoniuk, \pos{PoS (LHC08) 002} [{\tt arXiv:0810.2082}].   
\bibitem{hydjet} http://cern.ch/lokhtin/hydro/hydjet.html~.
\bibitem{Amelin:2006qe}  N.S.~Amelin et al., \emph{Phys. Rev.} {\bf C 74} (2006) 
064901.
\bibitem{Amelin:2007ic}  N.S.~Amelin et al., \emph{Phys. Rev.} {\bf C 77} (2008) 
014903.
\bibitem{root} R.~Brun and F.~Rademakers, \emph{Nucl. Instrum. Meth.} {\bf A 389} 
(1997) 81; (http://root.cern.ch). 
\bibitem{pyquen} http://cern.ch/lokhtin/pyquen~.
\bibitem{pythia} T.~Sjostrand, S.~Mrenna and P.~Skands, \emph{JHEP} {\bf 0605} (2006) 
026. 
\bibitem{Tywoniuk:2007xy} K.~Tywoniuk, I.C.~Arsene, L.~Bravina, A.B.~Kaidalov 
and E.~Zabrodin, \emph{Phys. Lett.} {\bf B 657} (2007) 170.
\bibitem{Renk-LHC09} T.~Renk, in this Proceedings 
\bibitem{Zapp-LHC09} K.~Zapp, in this Proceedings 
\bibitem{Mendez-LHC09} L.~Mendez, in this Proceedings
\bibitem{share} G.~Torrieri et al., \emph{Comput. Phys. Commun.} {\bf 167} 
(2005) 229. 
\bibitem{therminator} A.~Kisiel, T.~Taluc, W.~Broniowski and W.~Florkowski, 
\emph{Comput. Phys. Commun.} {\bf 174} (2006) 669.
\bibitem{Back:2002wb} B.B.~Back {\it et al.} [PHOBOS Collaboration], 
\emph{Phys. Rev. Lett.} {\bf 91} (2003) 052303. 
\bibitem{Abelev:2006jr} B.I.~Abelev {\it et al.} [STAR Collaboration], 
\emph{Phys. Rev. Lett.} {\bf 97} (2006) 152301.
\bibitem{Adams:2004bi} J.~Adams {\it et al.} [STAR Collaboration],
\emph{Phys. Rev.} {\bf C 72} (2005) 014904. 
\bibitem{eyyubova-LHC09} G.~Eyyubova, in this Proceedings
\bibitem{milosevic-LHC09} J.~Milosevic, in this Proceedings
\bibitem{Paic:2004zj} G.~Paic, P.K.~Skowronski and B.~Tomasik, 
\emph{Nucleonika} {\bf 49} (2004) S89 [{\tt nucl-th/0403007}].
\bibitem{pod83} M.I.~Podgoretskii, \emph{Sov. J. Nucl. Phys.} {\bf 37} (1983) 272.
\bibitem{Pratt84} S.~Pratt, \emph{Phys. Rev. Lett.} {\bf 53} (1984) 1219.
\bibitem{CF_STAR} J.~Adams {\it et al.} [STAR Collaboration], \emph{Phys. Rev.}  
{\bf C 71} (2005) 044906.

\end{thebibliography}
\end{document}